\documentclass[10pt, conference, compsocconf]{IEEEtran}

\usepackage{graphicx}
\usepackage[numbers]{natbib}
\usepackage{multirow}
\usepackage{tabularx}
\usepackage{threeparttable}
\usepackage{amssymb}
\usepackage{amsmath}
\usepackage{bm}
\usepackage[hyphens]{url}
\usepackage{color}

\newcommand{\mean}[1]{\overline{#1}}

\begin{document}

\title{Method for estimating cycle lengths from multidimensional time series:\\
	Test cases and application to a massive ``in silico'' dataset}

\author{
    \IEEEauthorblockN{N. Olspert\IEEEauthorrefmark{1}, 
    	M. J. K\"apyl\"a\IEEEauthorrefmark{1}\IEEEauthorrefmark{2} and
    	J. Pelt\IEEEauthorrefmark{3}}
    	\\
    \IEEEauthorblockA{\IEEEauthorrefmark{1}Department of Computer Science,
    	ReSoLVE Centre of Excellence, Aalto University,
    	PO Box 15400,\\FI-00076 Aalto, Finland,
    	Email: nigul.olspert@aalto.fi}
    	\\
    \IEEEauthorblockA{\IEEEauthorrefmark{2}
    	Max-Planck-Institut f\"ur Sonnensystemforschung,
    	Justus-von-Liebig-Weg 3, D-37077 G\"ottingen, Germany}
    	\\
    \IEEEauthorblockA{\IEEEauthorrefmark{3}Tartu Observatory,
    	61602 T\~{o}ravere, Estonia}
}

\maketitle

\begin{abstract}
Many real world systems exhibit cyclic behavior that is, for example, due to the nearly harmonic oscillations being perturbed by the strong fluctuations present in the regime of
significant non-linearities. For the investigation of such systems special techniques
relaxing the assumption to periodicity are required. In this paper, we present the
generalization of one of such techniques, namely the $D^2$ phase dispersion statistic, to
multidimensional datasets, especially suited for the analysis of the outputs from 
three-dimensional numerical simulations of the full magnetohydrodynamic equations. We 
present the motivation and need for the usage of such a method with simple test cases, and present an application to a solar-like semi-global numerical dynamo simulation covering 
nearly 150 magnetic cycles.
\end{abstract}


\IEEEpeerreviewmaketitle

\section{Introduction}
The analysis method discussed in this paper belongs to the group of phase dispersion 
minimization (PDM) methods first introduced by \cite{Lafler1965, Stellingwerf1978}. 
The $D^2$ statistic, in particular, was formulated in \cite{Pelt1983}. While
these methods have been widely used in period search from variable star light 
curves for many decades, they have a limitation to the time series with a stable period persistent over a long time span.
A modification to the $D^2$ statistic, that relaxes this condition, assuming that the period 
can slightly vary over time around a certain mean value, was introduced by employing a window function \cite{Pelt1983, Pelt1993}. We will refer to this mean oscillation time as the {\it cycle length.} 
For these kinds of cyclic time series, the $D^2$ statistic becomes more favorable over 
the other PDM as well as spectral analysis methods, e.g. Lomb-Scargle periodogram \cite{Lomb1976, Scargle1982}, as the spectra 
for the latter ones may be hard to interpret due to the emergence of sideband components as a result of the modulation of the periodic signal. 

Alternatively, there are methods belonging to a class of so called time-frequency or
multiresolution analysis. 
These methods are especially suitable for dealing with nonstationary data where the 
direct interpretation of the power spectrum is impossible. 
In Wavelet Transform (WT) method,
contracted and dilated versions of a single prototype function (called a wavelet) are used to analyze the
signal at different scales \cite{Vetterli1992}. 
In Empirical Mode Decomposition method, the signal is adaptively decomposed into basis functions or
modes which are derived from the data \cite{Huang1998}. To overcome the mode-mixing problem 
in the latter method, a noise-assisted approach was introduced, called Ensemble Empirical
Mode Decomposition (EEMD) \cite{Wu2009}. It allows for extracting true and physically meaningful modes from the data.
The multiresolution aspect achieved by both of the methods is easier to understand if we think of them as dyadic filter banks.
While these methods are valuable in tracking the local transitions and
discontinuities as well as long-term behaviour from time series,
they are limited to uniform sampling. 
Concerning EEMD, our previous experience 
has also shown that it is computationally demanding due to iterative envelope fitting 
and large ensemble size required by the noise-assisted approach. 

In contrast to WT and EEMD, the $D^2$ statistic cannot answer the question about 
the locality of the events in time series, but it addresses questions such as the existence of
cyclic behaviour and the stability of the cycle. 
Moreover, it is suitable for unevenly
sampled time series and most importantly, what constitutes the main topic of this paper,
it is easily generalizable to multidimensional time series such as the data sets being produced
to increasing extent and size by 
fully three-dimensional (3D)
numerical models of high complexity.

The aim of this paper is to present proof-of-concept cases for the necessity and usefulness of the multidimensional $D^2$ statistic, and highlight its power by analysing the 
solar-like semi-global 3D magnetoconvection simulation \cite{Kapyla2016a}, 
denoted as PENCIL-Millennium, exhibiting solar-like cycles with irregular behavior.

\section{Method}\label{method}
The direct application of the $D^2$ statistic is to estimate periods or lengths of the 
cycles from time series.
One of the most important benefits of the method
is the suitability for the data sets with irregular sampling, 
making it applicable especially for astronomical datasets where data gaps are more a rule than 
an exception. 
It has been widely used to study stellar rotation periods, the rotation of magnetic spot structures, and stellar magnetic cycles, two recent examples having been presented in
\cite{Olspert2015,Lindborg2013}.
The method could be applied to study satellite measurements (e.g. KEPLER, PLATO), although
these time series are too short for the investigations of stellar magnetic activity cycles, which are
our main focus. While virtually all observational time series are yet too short, numerical simulations provide data
that span over tens and even hundreds of cycles, and also provide a view inside the star. The 
inevitable consequence is the data becoming multidimensional, to which aspect we concentrate here.

The multidimensional generalization of the $D^2$ phase dispersion statistic can be written as
\begin{equation}\label{d2_def}
D^2(P,t_{\rm coh})\!=\!\frac{\sum\limits_{i=1}^{N-1} \sum\limits_{j=i+1}^N g(t_i, t_j, P, t_{\rm coh})||{\bm f}(t_i) - {\bm f}(t_j)||^2}{2\sigma^2\sum\limits_{i=1}^{N-1} \sum\limits_{j=i+1}^N g(t_i, t_j, P, t_{\rm coh})},
\end{equation}
where $N$ is the number of data points, ${\bm f}(t_i) \in \mathbb{R}^D$ is the $D$-dimensional vector of observed variables at time
moment $t_i$, $\sigma^2=\frac{1}{N(N-1)}\sum_{i,j>i}||{\bm f}(t_i) -
{\bm f}(t_j )||^2$ is the variance of the full time series,
$g(t_i,t_j,P,t_{\rm coh})$ is the selection function, which is
significantly greater than zero only when
\begin{eqnarray}
  t_j  - t_i  &\approx& kP,k =  \pm 1, \pm 2, \ldots, \quad \mbox{and}\\
\left| {t_j  - t_i } \right| &\lesssim& t_{\rm coh} = l_{\rm coh}P,
\end{eqnarray}
where $P$ is the trial period and $t_{\rm coh}$ is the so-called
coherence time, which is the measure of the width of the sliding time window
wherein the data points are taken into account by the statistic.
The number of trial periods fitting into this interval, $l_{\rm coh}=t_{\rm coh}/P$, is called a coherence length.
The only requirement for the applicability of the statistic is the existence of the suitable norm for the
vector of observed variables (in our analysis we used Euclidean norm).

We note that the physical meaning of the dimensions of the multivariate time series may
differ from one problem to another. 
Examples are, e.g., time series
of three components of a vector quantity, time series of scalar quantity measured at each
time moment in multiple points in space, and in the most general case a time series
of multiple vector quantities measured over a certain volume in space. 

The form of the selection function $g(t_i,t_j,P,t_{\rm coh})$ is a matter of preference 
depending on the dataset being analyzed. 
Computations are fast for a box-type 
function, defined as 
\begin{equation}\label{box_weight_fn}
g=\mathbb{I}(|t_i-t_j| < t_{\rm coh})\mathbb{I}({\rm frac}(\nu|t_i-t_j|) < \epsilon), 
\end{equation}
where $\mathbb{I}$ is the indicator function, $\nu=1/P$ and $\epsilon$ is 
maximum allowed phase separation (taken usually as 0.1).
This kind of selection function is well suitable for irregularly sampled data. 
For evenly sampled data a preferred form is a Gauss-cosine function, which smooths away possible artefacts in the spectrum, e.g. induced by the even sampling itself. This function can be defined e.g. as
\begin{equation}\label{gauss_cosine_weight_fn}
g=e^{-\ln 2 (t_i-t_j)^2/t_{\rm coh}^2}(1+2\cos(2\pi\nu(t_i-t_j))).
\end{equation}
The factor $\ln 2$ instead of $0.5$ 
appears as $t_{\rm coh}$ represents HWHM of the Gaussian.
The data from the MHD simulations is evenly sampled in the saturated regime of the dynamo
and therefore we use here the Gauss-cosine version of the selection function.

When $t_{\rm coh}$ is taken longer than or equal to the dataset
length only the phase selection terms remain in Eqs. (\ref{box_weight_fn}) and (\ref{gauss_cosine_weight_fn}). In the former case the $D^2$ statistic is closely related to 
the Stellingwerf statistic \citep{Stellingwerf1978} and in the latter case it coincides with the least-squares spectrum of the harmonic model as well as with residual power spectrum \cite{Pelt1983}.
The idea behind introducing the selection function depending on the time distance
between the data points comes from the fact that in the case of 
cyclic
signals the correlation between the data points, having same phase w.r.t. given trial period, is always lost after a
certain time lag regardless of the selected period. If the local correlation, however, is persistent throughout the data
set we can say that the time series has a mean cycle and be able to detect it.
Analyzing the patterns in the $D^2$ spectrum as function of the coherence
length also enables us, at least qualitatively, to describe what kind of
process we are dealing with.
The additional dependence on $t_{\rm coh}$ as well as the applicability to multidimensional data makes the $D^2$ statistic more general than most of the other widely used spectral estimation tools. The direct comparison with other methods is therefore impossible.

Next we shortly
describe the procedure of estimating the average cycle length from the spectrum and the significance of the minima obtained. 
If the signal is not exactly periodic then the spectra has usually a unimodal shape 
below and a multimodal\footnote{By multimodality we mean that at fixed coherence length the spectrum as a function of frequency only has multiple distinct minima.} shape above a certain coherence length (i.e. there is a split point).
If, when moving from that point towards shorter coherence lengths,
the position of the minimum stays constant, and the minimum does not get weaker,
we have obtained a suitable coherence length, which we
interpret as the average coherence length of the given cycle.
The significance of the found cycle lengths can be estimated using e.g.
randomization proposed in \cite{Nemec1985}, in which case the null hypothesis corresponds
to white noise. In many cases it is obvious that the data is resembling red rather
than white noise, so the usage of this assumption leads to the overestimation of the cycle length significances. However, to correctly
set a hypothesis, we would first need to estimate the red noise model from the data. 
In spectral
analysis this is done e.g. by fitting the power law to the periodogram \citep{Vaughan2005}. 
Here, we adopt a more simplistic approach.
As in our case the definition of the $D^2$ statistic assumes stationarity
\footnote{The selection function $g$ does not depend on time moments directly, but only on time differences}, we can use
Bootstrap resampling of time-observation pairs, which is 
equivalent to drawing samples from the distribution of square differences of $\mathbf{f}$ in Eq.~(\ref{d2_def}).
This procedure enables us to obtain error estimates for the 
mean cycle lengths.

We finish this section with a few notes concerning the computational complexity of
the $D^2$ statistic. Written as Eq.~(\ref{d2_def}), it has the computational complexity of
$O(N^2 \times D)$ per each trial period $P_i$, $i = 1 \ldots N_{\rm P}$ and coherence length
${l_{\rm coh}}_j$, $j = 1 \ldots N_{\rm l_{\rm coh}}$,
the overall complexity of the calculation of a spectrum amounting to $O(N^2 \times D \times N_{\rm P} \times N_{\rm l_{\rm coh}})$. However, as a first
step we can calculate the partial sums 
\begin{equation}
S_k=\sum\limits_{(k-1) \Delta t<|t_i-t_j|<k\Delta t}||{\bm f}(t_i) - {\bm f}(t_j )||^2,
\end{equation}
where $k = 1 \ldots K$, $K={\rm ceil}((t_N-t_1)/\Delta t)$ and $\Delta t$ is selected small enough compared 
to the lowest trial period
$P_{\rm min}$ in the search range. Secondly, if the average number of data points
falling into an effective window of the longest coherence time is $M$, then the overall complexity reduces to
$O(N \times M \times D + K \times N_{\rm P} \times N_{\rm l_{\rm coh}})$.
Further optimizations can be achieved by expressing the statistic in a form
of trigonometric sums and taking advantage of FFT \citep{Pelt1983}, but this is
out of the scope of the current study. 

Regardless of the achieved performance gain by using the 
scheme above, parallelization is inevitably needed when used on a dataset with a considerable size. 
For example, at the moment of writing this paper the PENCIL-Millennium
dataset contained $N\approx20000$ data points with
dimensionality $D=3\times128\times256=98304$. In a limiting case of $M=N$, the first term in the above
complexity formula is around $4\times10^{13}$, the second term being negligible.
We performed all the calculations 
using CSC supercluster Taito, where we used eight nodes with eight CPU's each running at 2.6 GHz. Depending on the subdomain of the data being analysed and the period search range selected, a single computation lasted from tens of minutes to close to two hours without bootstrap resampling. 

\section{Applications}\label{applications}
\subsection{Experiments with simple test cases}\label{tests}
The evident benefit of the multidimensional statistic is
the possibility to abandon the piece-wise approach, that is, to work with 
one-dimensional
data sets separately. 
As we will illustrate with the following simple examples, the piece-wise approach can even lead to the non-detection of a cycle, a problem that can be remedied by using a multidimensional statistic.

\begin{figure}
\centering
\includegraphics[width=0.25\textwidth]{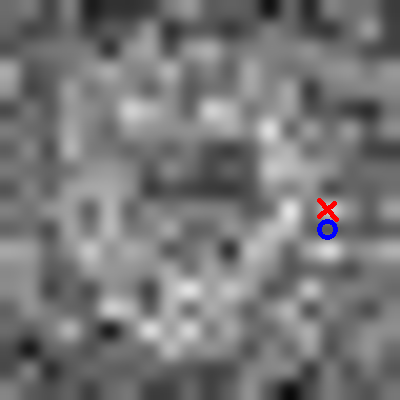}
\caption{Average path of the test particle. White indicates higher visiting frequency, black lower. Red cross and blue circle show the points of sampling used in
one-dimensional analysis.}\label{exp1_path}
\end{figure}

\begin{figure}
\begin{center}
\includegraphics[width=0.4\textwidth]{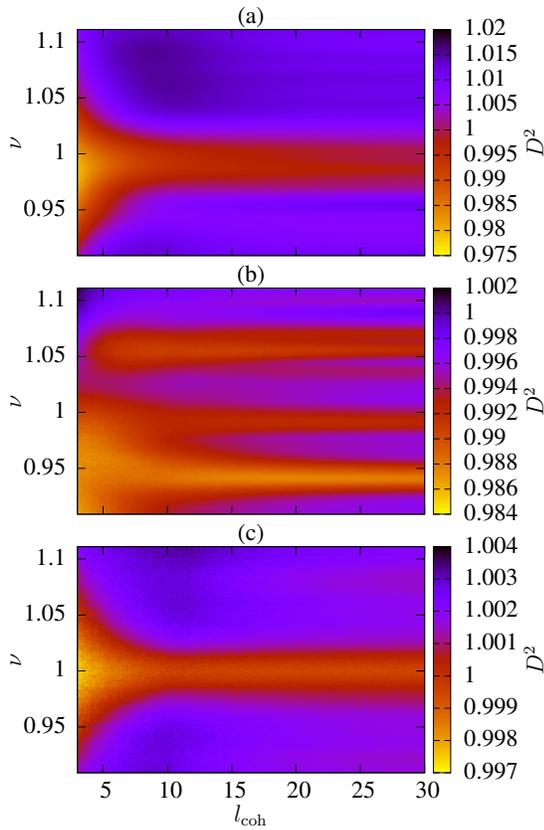}
\caption{$D^2$ spectrum of the rotating particle on a trajectory influenced by random noise. Analysis is done using (a) point of high visiting frequency, (b) random point, (c) full grid.}\label{exp1_d2}
\end{center}
\end{figure}

In the first example we have a test ``particle'' moving periodically around the center 
on a trajectory that is influenced by random noise. 
The simulation domain is a square of the dimensions $20\times20$ and
the average rotation period is taken equal to one time unit. 
The average trajectory of such a ``particle'' can be seen in 
Fig.~\ref{exp1_path}. As the fluctuations push and pull the particle around,
it does not form nice closed loops, but rather some dispersed cycles. In this
simple case we can sample the space 
approximately at the points of the most probable visits and carry out the
analysis in one
dimension. The results of one of the samples of high visiting frequency (marked with red cross
on Fig.~\ref{exp1_path}) is
shown on Fig.~\ref{exp1_d2}(a). 
As we see the spectrum is biased towards the lower frequencies as the
``particle'' still does not pass the selected point on every rotation.
Moreover, selecting the optimal point from simulation space for the
analysis becomes harder or 
infeasible if the number of dimensions is more than two. Picking spatial samples
randomly out of the dataset is neither a good idea, 
because even in our simple case by offsetting the sample by one pixel (marked with blue circle
on Fig.~\ref{exp1_path})
we obtain meaningless results as seen on Fig.~\ref{exp1_d2}(b). 
The full two-dimensional $D^2$ analysis is shown 
on panel (c) of the same figure and yields correctly the expected period of one.

\begin{figure}
\begin{center}
\includegraphics[width=0.4\textwidth]{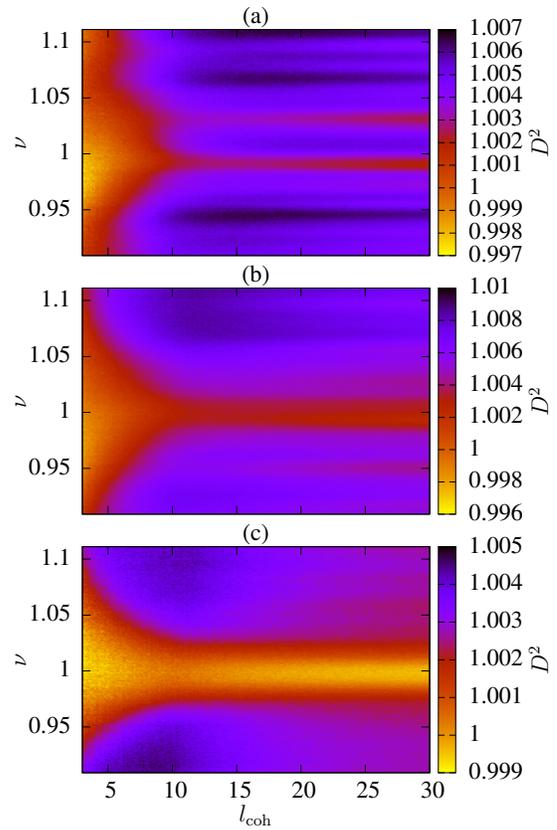}
\caption{$D^2$ spectrum of the weak oscillations in the noisy environment. Analysis is done using (a) 4 grid points, (b) 16 grid points and (c) full grid.}\label{exp2_d2}
\end{center}
\end{figure}

As a second example we again consider a square sampled with a grid of $20\times20$
points, but now the whole scalar field inside the volume is oscillating with
the same period. 
The average period is again taken equal to one time unit. 
The amplitude of the oscillation is taken five times smaller than the 
added white
Gaussian noise and the noise is not correlated from point to point.
If all these facts were known a priori, we could sample the full grid point by point and sum up the time
series to eliminate the noise. Then we could proceed with one-dimensional analysis. 
For this technique to work, however,
the phases of the oscillations must be coherent at different points in the simulation space. 
By using the $D^2$ statistic over the full volume, we do not have this kind of a restriction,
neither do we need to preprocess the data to reduce the dimensionality. The results of this
experiment are shown in Fig.~\ref{exp2_d2}. On the panel (a) we have made $D^2$ analysis for
the subregion consisting of four grid points, on the panel (b) for 16 points and on the panel (c)
for the full grid. As we see the spectrum converges around the correct oscillation
period as we increase the number of points used in the analysis.

\subsection{Periodic time series with varying signal-to-noise ratio}
Next, we address some issues concerning the interpretation of the spectra, that are important to 
consider before drawing any conclusions.
For this purpose we have generated some additional artificial time series and calculated the
corresponding $D^2$ spectra. The aspects discussed in this and the following two 
subsections are relevant regardless
of the dimensionality of the data.

\begin{figure*}
\begin{center}
\includegraphics[width=0.77\textwidth]{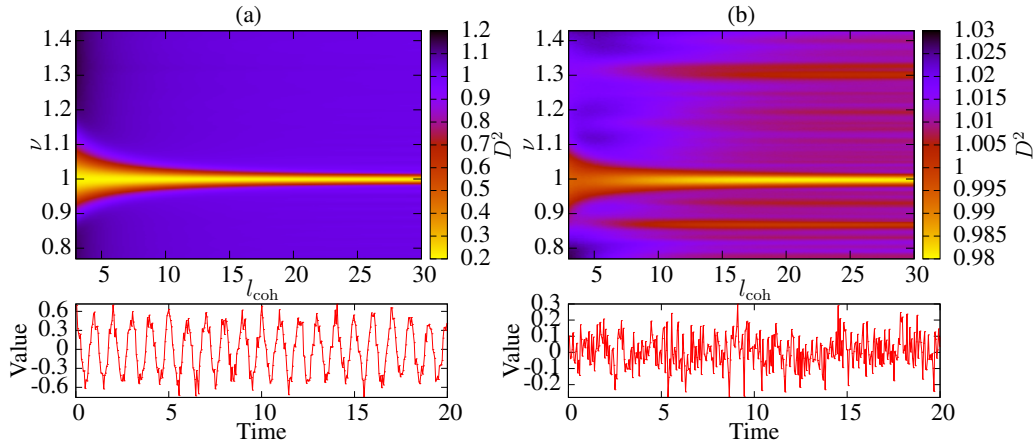}
\caption{Patterns of periodic signal. (a) Signal with high S/N ratio. (b) Signal
with low S/N ratio. On the bottom: excerpts from the corresponding time series.}
\label{pattern_periodic}
\end{center}
\end{figure*}

As a first example we have taken a periodic time series with both a high and a low S/N
ratios (5 and 0.2 respectively), 
the $D^2$ results being shown in Fig.~\ref{pattern_periodic}(a) and (b), respectively.
The following important features can be observed from these plots.
Firstly, in the case of a high S/N ratio the peak of the spectrum has a constant
amplitude regardless of the coherence length. Secondly, for periodic signals the
spectrum does not split, but stays unimodal at constant frequency. Thirdly, for the low S/N ratio case we
see that the form of the spectrum is similar, only with the difference that the
minimum of it gets stronger as we move from shorter coherence lengths to longer ones.
This means that the noisier the data is the greater number of periods we need to
include into the analysis to get statistically significant estimates and
even then the minimum is very weak compared to the high S/N ratio case. 
The fourth observation from these plots is that in the case of noisy data, 
due to the finite length of the time series, artificial spectral lines appear.

\begin{figure*}
\begin{center}
\includegraphics[width=0.77\textwidth]{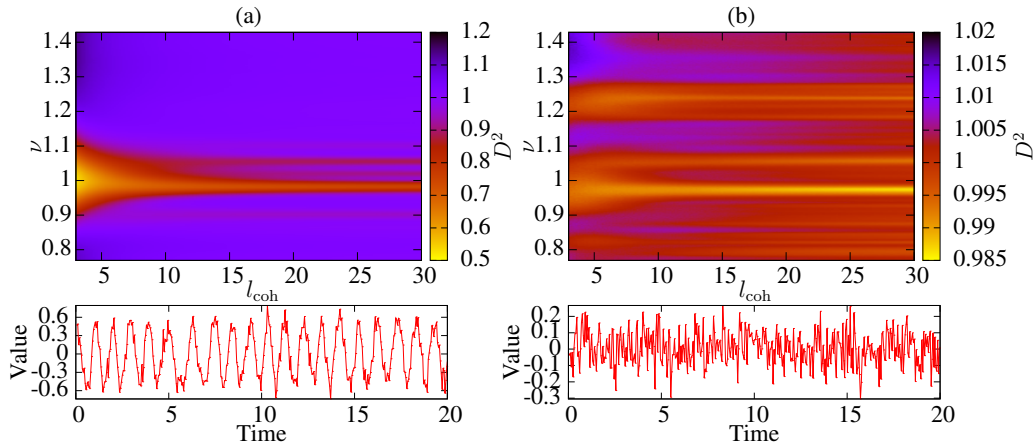}
\caption{Patterns of cyclic signal. (a) Signal with high S/N ratio. (b) Signal with low S/N ratio. On the bottom: excerpts from the corresponding time series.}
\label{pattern_cyclic}
\end{center}
\end{figure*}

\subsection{Cyclic time series with varying signal-to-noise ratio}
In the second example we consider a signal, which is no longer periodic, but
the period is slightly changing around a certain mean value over time, i.e. the
signal is now cyclic. Again we
have considered both high and low S/N ratio cases. The corresponding results are
shown in Fig.~\ref{pattern_cyclic}(a) and (b), respectively. In the case of low noise
we see a clear difference in comparison to the periodic case, Fig.~\ref{pattern_periodic}(a), such that 
the main minimum is observed to loose its power as a function of coherence length,
and shift in frequency. The spectrum also becomes multimodal starting from a certain coherence length. With this kind
of splitting the separate spectral lines at high coherence lengths are usually
shifted w.r.t. and scattered around the main minimum at low coherence length, 
the latter one representing the true mean cycle length of the time series.
We note that at high values of $l_{\rm coh}$ the spectrum is 
nearly identical to Fourier power spectrum. However as we saw in this example, neither the strongest nor any other peak of this spectrum corresponds to the true cycle length. Thus one of the key benefits of $D^2$ statistic compared to the other methods is the possibility 
to detect cycle lengths even for strictly non periodic signals.
As in the periodic case, the spectrum for the noisy data shows many minima that become
enhanced as function of the coherence length, but already at smaller values of the coherence length than with the periodic signal. In this case, even the deepest minimum is offset
from the true value towards longer cycle length. We do not observe any convergence to 
a unimodal shape either.
We conclude that if the cyclic time series has a low S/N ratio, it is not a trivial task to deduce
the cycle length from the $D^2$ spectrum. 

\begin{figure*}
\begin{center}
\includegraphics[width=0.76\textwidth]{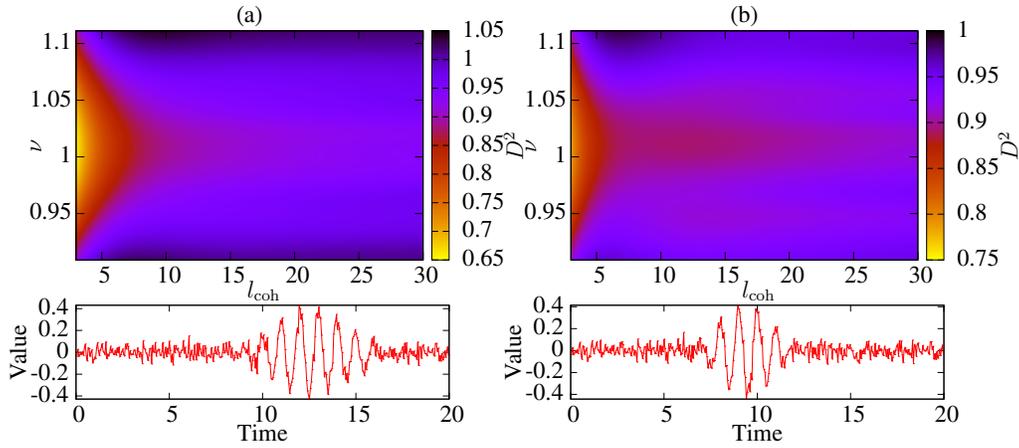}
\caption{Patterns of temporary signal with longer duration (a) and 
shorter duration (b). On the bottom: excerpts from the corresponding time series.}
\label{pattern_temporary}
\end{center}
\end{figure*}

As the last example we consider a time series where the S/N ratio is high, but the
cycle appears only temporarily. In a multidimensional case this can correspond
to a slowly migrating oscillating subregion in the domain. 
In Fig.~\ref{pattern_temporary} we have plotted two cases where
on the panel (a) the cycle lasts approximately for six time units and on
the panel (b) for four time units, respectively,
the overall dataset length being 50 time units in both
cases. The main observation from these results is that the shorter the duration
of the cyclic signal is compared to the total length of the time series, 
the shorter is the cut-off coherence length starting from which the 
spectral power drops rather abruptly.

\begin{figure}
\begin{center}
\includegraphics[width=0.39\textwidth]{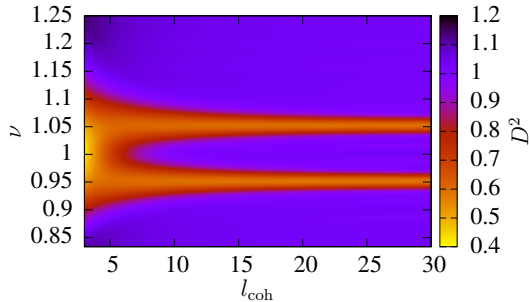}
\caption{$D^2$ spectrum of time series with multiple cycles at different subregions.}\label{pattern_2_periods}
\end{center}
\end{figure}

We conclude this section by describing a small caveat. In the above examples we have used an assumption that in the subspace of the variables there is
only a single cyclic process. If for instance there are two separate regions with different cycle
lengths then interpretation of the spectrum must be taken with caution. If the cycle lengths in
these two regions differ significantly, then in the resulting spectrum two distinct minima
appear. If, however the cycle lengths are quite close then at coherence lengths below certain
threshold the minima will merge together as shown on Fig.~\ref{pattern_2_periods}. In this particular case we have two
fully periodic processes with periods slightly above and below one time unit, 
but a combined minimum appears at
very small coherence lengths. Exactly similar pattern occurs when
the time series consists of two processes with distinct cycle lengths, one of them being active
during the first and the other during the second half of the time series. 
To confirm or disprove
the given spectrum being a manifestation of such processes the analysis should be repeated for
subspace of variables and/or for smaller time windows.

As the last remark we note that, like with any other period estimation method, due to data sampling, aliases appear in the $D^2$ spectrum. In astronomical time series the aliases are usually caused by Earth's rotation or seasonal patterns in the observations and may result in detection of spurious periods. Techniques for eliminating these periods
can be found from \cite{Jetsu2000} or \cite{Tanner1948}.

\subsection{Multicyclic time series}\label{multicyclic}
The $D^2$ statistic defined by Eq. (\ref{d2_def}) works reasonably well on the time
series representing a process with a single cycle or period. If there are more
cycles in the data then each one of them essentially acts as noise on the phase
diagram constructed for the others. In the case of data sets with high S/N ratio this is
not a problem, but in other cases the weak minima in $D^2$ spectra may be hard or impossible 
to detect. One possible way to take this effect into 
account would be to generalize the statistic so that it would be dependent on multiple periods
\citep{Pelt1993}. This approach, however, works only in the case when the shorter
cycle has average coherence extending at least over multiple longer cycles. 
Another disadvantage of the method is that a significantly longer time series would be
needed as the number of data point pairs, for which the phase proximity w.r.t. two trial periods is small, is reduced on average by a factor of $\epsilon$, compared to
the single period case.
An alternative way to overcome this problem is to first estimate the longer cycle length with
the $D^2$ statistic, then use regression, e.g. Carrier fit \citep{Pelt2011}, to remove this
cycle from the data and 
continue with estimating the shorter cycle. 
As for multidimensional time series the need for
regression at each grid point leads to a significant increase in computational time,
this solution might be impracticable.

In the current study, as we work with evenly sampled data, we take a simpler approach and
subtract moving average of suitable width
from data to eliminate the effect of longer cycles when estimating the shorter ones. This is
approximately equivalent to high-pass filtering of the signal. 
A slight modification to the square difference term in Eq. (\ref{d2_def}) is needed, 
where instead of
${\bm f}(t_i) - {\bm f}(t_j)$ we have ${\bm f}(t_i) - \mean{{\bm f}}(t_i) - {\bm f}(t_j) +
\mean{{\bm f}}(t_j)$, where $\mean{{\bm f}}(t_i)=1/n\sum_{i-n}^{i+n}{\bm f}(t_i)$ is the moving
average of the input vector at time moment $t_i$. 
The smoothing width $n$ can be adjusted according to the upper limit of the period search range.

\subsection{Data from a 3D MHD model}\label{data}
Only very recently has it become possible to model solar and stellar cycles with numerical models, that solve for the magnetohydrodynamic equations in spherical geometry, and obtain solutions that resemble the observed behavior of the magnetic cycle (roughly 22-year magnetic cycle with strong irregularities). The advantage of numerical models over 
observations is that they provide a fully 3D view of the physical processes 
throughout the convection zone, and reveal the working mechanisms of the dynamo process generating and sustaining the 
magnetic field. The ``silico'' data sets are multidimensional, and provide the most direct 
application to the method presented in this paper. In this study, we analyze data from the PENCIL-Millennium simulation, which has currently been integrated over 
150 solar-like cycles. 
Analysis of the simulation data over the first 80 magnetic cycles and 
details of the model were presented in \cite{Kapyla2016a},
where we used EEMD by sampling the data at different locations of the simulation domain. We were
able to identify three different cycles, but the analysis remained indecisive for the shortest cycle,
for which a mean cycle length could not be determined. With a small modification to 
the $D^2$ statistic introduced in Sect.~\ref{multicyclic}, this becomes possible.

As the solution obtained is axisymmetric, i.e. does not vary as function of the azimuthal coordinate,
we use azimuthally averaged data in our analysis. The data, therefore, represents a time series of 42 physical 
quantities measured on $128\times256$ spatial grid. As observable solar and stellar activity tracers 
(starspots, CMEs etc.) all have a magnetic origin and occur in the regions of strong magnetic field,
we focus our analysis on the components of the mean magnetic field vector:
radial - $\mean{B}_r$, latitudinal - $\mean{B}_\theta$ and toroidal - $\mean{B}_\phi$. 

\begin{figure*}
	\begin{center}
		\includegraphics[width=0.92\textwidth]{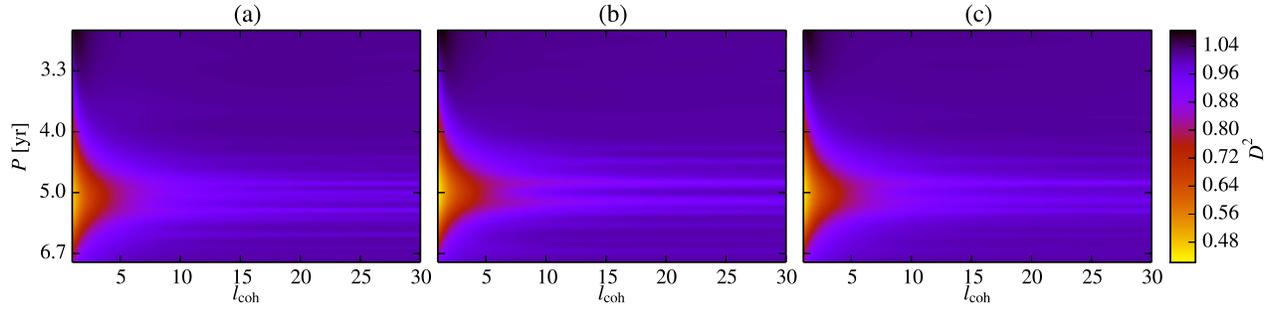}
		\caption{$D^2$ spectra of $\mean{B}_\theta$ revealing the five year cycle. (a) Northern
		 hemisphere, (b) southern hemisphere, (c) full meridional plane.}\label{res_5y_cycle}
	\end{center}
\end{figure*}

We started by analyzing the full meridional plane in a period range from 0.1 to 146 years.
The upper limit for the period search range was set with the requirement that at least five
full cycles would be covered by the dataset. After the pilot search we detected one
minimum around five years, confirming the main result from EEMD analysis in
\cite{Kapyla2016a}. 
The results, after refining the period search range, showed that the spectra for all
components of the magnetic field vector look similar, deviating only
in the depth of the minima. The cycle appears strongest for $\mean{B}_\theta$ and the
corresponding $D^2$ spectrum is shown on Fig.~\ref{res_5y_cycle}(c).
We subsequently repeated the analysis separately for both hemispheres, 
the results being shown on Fig.~\ref{res_5y_cycle}(a) and (b). 
As we see the mean cycle lengths are roughly equal 
(but not exactly as will be later shown) on both hemispheres and
as expected, the global spectrum on panel (c) is just an average of
the spectra for north and south hemispheres taken separately.
The spectrum for the northern hemisphere has a split point at slightly shorter coherence
length indicating a greater overall cycle length variation compared to the southern hemisphere.

We further divided the meridional plane into two pieces latitudinally in both hemispheres
and into three pieces radially. Based on the local spectra calculated for these regions
separately we observed that for $\mean{B}_r$ the cycle is present throughout the full domain,
while for the other components the cycle is not detected in the bottom
layer from equator towards mid-latitude regions on both hemispheres, where the spectra show similar
patterns as on Fig.~\ref{pattern_temporary}. 
This might be an indication of the cycle not being existent for the full duration, but intermittently switching on or off. 
More detailed analysis of the time series from these regions is needed to answer this question.

\begin{figure*}
\begin{center}
\includegraphics[width=0.67\textwidth]{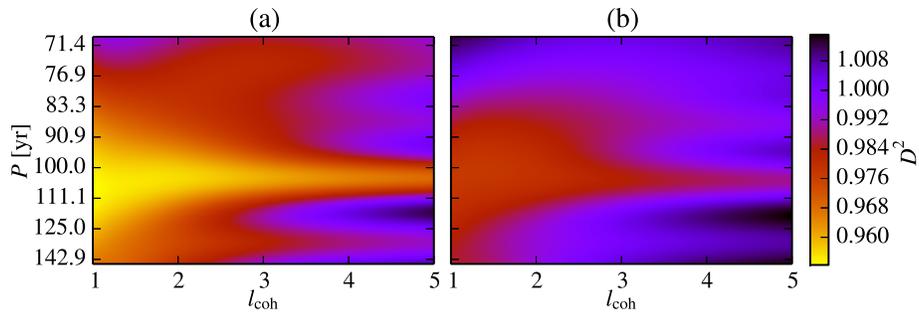}
\caption{Long cycle in the bottom of the convection zone. Shown are $D^2$ spectra for $\mean{B}_\phi$ for the northern hemisphere (a) and for the southern hemisphere (b).}\label{res_100y_cycle}
\end{center}
\end{figure*}

From the global spectrum 
we did not detect any additional minima, but the local spectra for the bottom quarter in
both hemispheres revealed a long cycle around 100 years. The corresponding spectra for $\mean{B}_\phi$ are depicted on Fig.~\ref{res_100y_cycle}. The spectra for $\mean{B}_r$ and
$\mean{B}_\theta$ revealed similar patterns, but the minima were slightly weaker. We further observed
that for $\mean{B}_r$ the cycle was stronger on the southern hemisphere while for
$\mean{B}_\phi$ and $\mean{B}_\theta$ on the northern hemisphere (difference in the case of $\mean{B}_\phi$ is clearly seen on Fig.~\ref{res_100y_cycle}).
More detailed analysis showed that the patterns in the spectra were
varying in the different latitudinal regions,
but the dataset is still too short to draw any definite conclusions.
Yet another important thing worth noticing is that the minima are weak in comparison to one 
corresponding to the five year cycle. 
Only for the toroidal component in the northern hemisphere the difference 
from general noise level is more than 5~\%. On one hand no strong minima were expected,
because the five year cycle already explains approximately 40--60~\% of the variance in the data,
on the other hand due to the weakness of the long cycle, plenty of variance remains still unexplained. 

\begin{figure}
\includegraphics[width=0.41\textwidth]{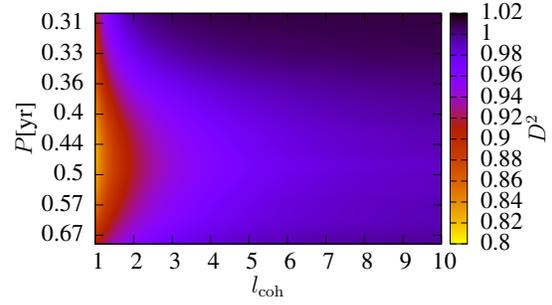}
\caption{Short cycle detected using $D^2$ statistic over full domain of $\mean{B}_\phi$.}\label{res_short_cycle}
\end{figure}

Without additional filtering no more cycles could be detected from the data, because the
amplitudes of the two detected ones shadow the other possible cycles.
We continued our search with the modified statistic as introduced in \ref{multicyclic}. 
In the high-frequency end we detected a cycle near 0.5 years. 
This cycle, covering approximately 20~\% of the variance, was
persistent in the dataset regardless of the selected subdomain or vector component. An example
of the spectrum for $\mean{B}_\phi$ over the full region is given on Fig.~\ref{res_short_cycle}. 
The spectra for the other components and subregions were very similar. As can be seen from the plot,
this cycle is coherent maximally for about two cycles, which essentially means that on average
only two neighboring cycles have a roughly matching cycle length. 
We also note that the spectrum closely resembles the one
seen on the right panel of Fig.~\ref{pattern_temporary}. 
This suggests that the cycle may be
not persistent throughout the full data set.
These two possible scenarios can also explain why the given cycle was not detected using EEMD
in \cite{Kapyla2016a}.

\begin{figure*}
\begin{center}
\includegraphics[width=0.67\textwidth]{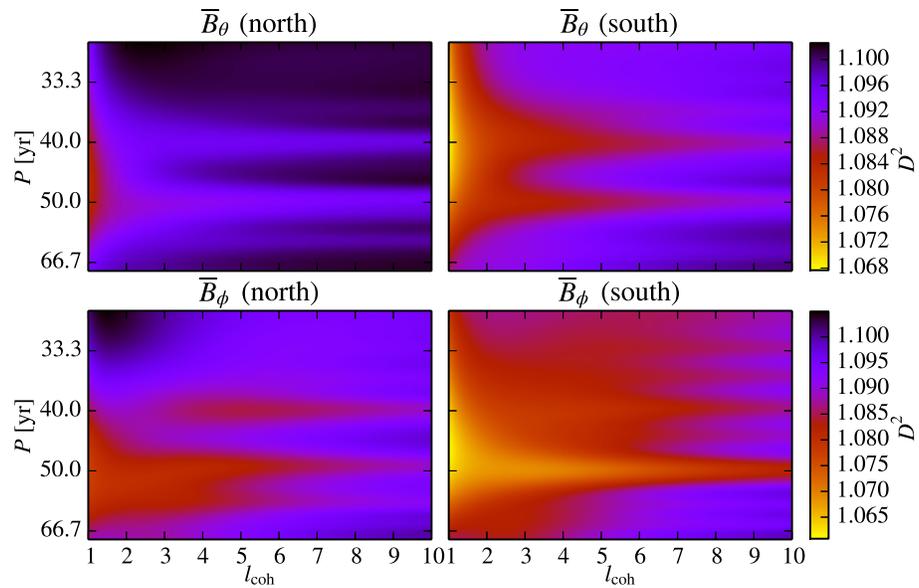}
\caption{$D^2$ spectra showing the 50 year cycle for $\mean{B}_\theta$ and $\mean{B}_\phi$. For $\mean{B}_r$ the cycle was seen only in the bottom of the convection zone.}\label{res_50y_cycle}
\end{center}
\end{figure*}

Using filtering we detected yet another cycle around 50 years,
which is so far the weakest one, explaining less that 3~\% of the variance.
For magnetic field component $\mean{B}_r$,
this cycle was prominent only in the bottom of the convection zone, while for the other components 
in the whole convection zone. 
In Fig.~\ref{res_50y_cycle} we have plotted the results for $\mean{B}_\theta$
and $\mean{B}_\phi$.
It is interesting to note that this cycle is stronger on the
southern than northern hemisphere, while component-wise the cycle is most prominent
in $\mean{B}_\phi$, and the weakest in $\mean{B}_r$ (not shown on the figure). 
These results somewhat diverge from the results seen in \cite{Kapyla2016a} using EEMD.
In the latter, cycle around 50 years could only be detected for $\mean{B}_\phi$ in the 
bottom of the convection
zone.
We conclude that due to the noise-assisted approach in EEMD, to be able to detect
weaker modes, considerably larger ensemble size would be needed than was used in the
aforementioned study.

The estimated mean cycle lengths with their 90~\% confidence intervals are gathered into
Table~\ref{table1}. 
We have used \textit{italic} font to indicate that the given cycle is present only
in the bottom of the convection zone. 
An immediate observation becoming evident from the table is the fact that the cycle lengths
for the northern hemisphere are slightly longer than for the southern hemisphere,
the only exception being cycle I in which case the cycle lengths exactly
match on both hemispheres.

As we see from the
Figs.~\ref{res_5y_cycle}, \ref{res_short_cycle} and \ref{res_50y_cycle} the coherence
lengths for all the cycles are very low -- shortest for the cycles I and III, being less than two
and longest for cycle II, maximally around five.
For cycle IV we cannot reliably determine the coherence length, as the spectra do not fully satisfy the above-mentioned criteria.
In some of the $D^2$ spectra (e.g. the long period range for $\mean{B}_r$) we saw a 
pattern similar to that seen on the right panels of
figures \ref{pattern_periodic} and \ref{pattern_cyclic} -- the minimum gets weaker towards the
lower coherence lengths. This is an indication that the noise is dominating the signal 
and caution is needed to avoid giving biased cycle estimates. 
As in all of those cases the unimodal minima 
around $l_{\rm coh}=1$ were clearly visible, we were nevertheless
able to estimate the cycle lengths from $D^2$ spectra.

\begin{table*}[t!]
\centering
\caption[]{Mean cycle length estimates}\label{table1}
\begin{tabular}{ccccccc}
\hline
\hline
\noalign{\smallskip}
\multirow{2}{*}{Cycle no} & \multicolumn{2}{c}{$B_r$} & \multicolumn{2}{c}{$B_{\theta}$} & \multicolumn{2}{c}{$B_{\phi}$} \\ 
& N & S & N & S & N & S \\
\hline
I & $0.47 \pm 0.01$ & $0.47 \pm 0.03$ & $0.48 \pm 0.01$ & $0.48 \pm 0.02$ & $0.46 \pm 0.02$ & $0.46 \pm 0.01$ \\
II & $5.12 \pm 0.06$ & $4.98 \pm 0.04$ & $5.13 \pm 0.05$ & $4.98 \pm 0.04$ & $5.17 \pm 0.05$ & $5.02 \pm 0.05$ \\
III & $\mathit{49.2 \pm 2.5}$ & $\mathit{43.0 \pm 1.5}$ & $46.2 \pm 1.1$ & $40.2 \pm 1.5$ & $50.8 \pm 1.6$ & $46.0 \pm 1.6$ \\
IV & $\mathit{108.4 \pm 5.3}$ & $\mathit{105.1 \pm 3.8}$ & $\mathit{108.0 \pm 3.5}$ & $\mathit{106.0 \pm 3.4}$ & $\mathit{107.5 \pm 3.7}$ & $\mathit{104.1 \pm 2.4}$ \\
\end{tabular}
    \begin{tablenotes}
      \small
      \item Notes: All the cycle length estimates are given in years. The numbers in \textit{italic} represent cycles appearing only
      in the bottom of the convection zone, otherwise the cycle exists in
      the full hemisphere. The error estimates correspond to 90~\% confidence intervals.
    \end{tablenotes}
\end{table*}

\section{Conclusions}\label{conc}
The $D^2$ statistic is not yet a fully developed and widely used method, thus its applicability and limitations
are still to be explored. In the given study we investigated the capabilities of the method,
generalized to multiple dimensions,
with the help of several artificially built test cases as well as a massive dataset from the
PENCIL-Millennium simulation. First we showed how different types of data sets lead to
different patterns in the spectrum and why multidimensionality aspect of the statistic 
is crucial for correctly determining the cycle length. 
Other important aspect supporting the usage of a multidimensional statistic is the 
possibility to gradually pinpoint the region of interest
from the data. With enough computational resources one could first start with global
analysis and, if hints of cyclic behaviour are seen, continue by ``zooming'' into the data
to fine-tune the results. 
The performance of the algorithm could be further improved by utilizing FFT as
the error estimation using Bootstrap sampling significantly increases the number of runs
needed.

From the results of PENCIL-Millennium analysis we point out the following findings. 
The strongest cycle around five years explains most of the variance in the data, but we confirm the
presence of other cycles: There is a short cycle about half a year length, invisible in an earlier attempt using EEMD - in this paper we pin-pointed on the reason for the non-detection, which is namely due to the extreme incoherency of this cycle. We also confirm the earlier finding of very long cycles of 50 and 100 years in the data, concentrated in the bottom of the convection zone, but the stability of these cycles cannot be answered due to time series still being too short.

One intriguing new aspect revealed by our analysis is the systematic difference between the cycle lengths in the north vs. south. It is well known that the dynamo solutions can exhibit significant hemispheric asymmetries - in the extreme case the dynamo cycle being visible only in one hemisphere alone, see e.g. the wedge simulation of turbulence magnetoconvection by \cite{Kapyla2010}. In the 
context of solar activity tracers, however, different behavior, e.g. different rotation periods, on the northern and the southern
hemispheres have, however, been related to weak non-axisymmetric activity nests, see e.g. \cite{Bai2003}. It is not ruled out that weak non-axisymmetric modes could be excited also in the PENCIL-Millennium, even though the wedge-assumption\footnote{We 
solve only a part of the full sphere, called as wedge, disregarding the poles and 
including only one quarter of the azimuthal extent.} is used in the azimuthal direction. The excitation of non-axisymmetric dynamo modes and azimuthal dynamo waves have already been reported in similar runs covering the full azimuthal extent, see e.g. \cite{Cole2014}. Investigating this issue further is, however, out of the scope of this study.

\newcommand{\etal}{et al.}

\section*{Acknowledgments}
This work has been supported by the Academy of Finland Centre of
Excellence ReSoLVE (NO, MJK, JP). The work of JP has also been supported
by Estonian Research Council (Grant IUT40-1).

\Urlmuskip=0mu plus 1mu\relax

\bibliographystyle{IEEEtran}
\bibliography{IEEEabrv,AAabrv,d2_multidim}

\end{document}